\input harvmac
\input amssym
\input epsf

\let\includefigures=\iftrue
%
%
%
%
%
\includefigures
\message{If you do not have epsf.tex (to include figures),}
\message{change the option at the top of the tex file.}
\def\figin{\epsfcheck\figin}\def\figins{\epsfcheck\figins}
\def\epsfcheck{\ifx\epsfbox\UnDeFiNeD
\message{(NO epsf.tex, FIGURES WILL BE IGNORED)}
\gdef\figin##1{\vskip2in}\gdef\figins##1{\hskip.5in}
\else\message{(FIGURES WILL BE INCLUDED)}%
\gdef\figin##1{##1}\gdef\figins##1{##1}\fi}
\def\DefWarn#1{}
\def\N{{\cal N}}
\def\figinsert{\goodbreak\midinsert}
\def\ifig#1#2#3{\DefWarn#1\xdef#1{fig.~\the\figno}
\writedef{#1\leftbracket fig.\noexpand~\the\figno}%
\figinsert\figin{\centerline{#3}}\medskip\centerline{\vbox{\baselineskip12pt
\advance\hsize by -1truein\noindent\footnotefont{\bf
Fig.~\the\figno:} #2}}
\bigskip\endinsert\global\advance\figno by1}
\else
\def\ifig#1#2#3{\xdef#1{fig.~\the\figno}
\writedef{#1\leftbracket fig.\noexpand~\the\figno}%
\global\advance\figno by1} \fi

\def\det{{\rm det}}


\def\IL{\relax{\rm I\kern-.18em L}}
\def\IH{\relax{\rm I\kern-.18em H}}
\def\IR{\relax{\rm I\kern-.18em R}}
\def\IC{\relax\hbox{$\inbar\kern-.3em{\rm C}$}}
\def\IZ{\relax\ifmmode\mathchoice
{\hbox{\cmss Z\kern-.4em Z}}{\hbox{\cmss Z\kern-.4em Z}}
{\lower.9pt\hbox{\cmsss Z\kern-.4em Z}} {\lower1.2pt\hbox{\cmsss
Z\kern-.4em Z}}\else{\cmss Z\kern-.4em Z}\fi}

\def\CZ {{\cal Z}}

\def\CC {{\cal C}}


\def\CZ {{\cal Z }}

\def\det{{\rm det}}
\def\Tr{{\rm Tr}}

\font\manual=manfnt \def\dbend{\lower3.5pt\hbox{\manual\char127}}

\def\IZ{\relax\ifmmode\mathchoice
{\hbox{\cmss Z\kern-.4em Z}}{\hbox{\cmss Z\kern-.4em Z}}
{\lower.9pt\hbox{\cmsss Z\kern-.4em Z}} {\lower1.2pt\hbox{\cmsss
Z\kern-.4em Z}}\else{\cmss Z\kern-.4em Z}\fi}

\def\lfm#1{\medskip\noindent\item{#1}}

\def\rt2{\sqrt{2}}
\def\irt2{{1\over\sqrt{2}}}

\def\slashchar#1{\setbox0=\hbox{$#1$}           
   \dimen0=\wd0                                 
   \setbox1=\hbox{/} \dimen1=\wd1               
   \ifdim\dimen0>\dimen1                        
      \rlap{\hbox to \dimen0{\hfil/\hfil}}      
      #1                                        
   \else                                        
      \rlap{\hbox to \dimen1{\hfil$#1$\hfil}}   
      /                                         
   \fi}

\lref\SeibergAT{ N.~Seiberg and D.~Shih, ``Minimal string
theory,'' {\tt hep-th/0409306}.
}

\lref\FateevIK{ V.~Fateev, A.~B.~Zamolodchikov and
A.~B.~Zamolodchikov, ``Boundary Liouville field theory. I:
Boundary state and boundary  two-point function,''
{\tt hep-th/0001012}.
}

\lref\TeschnerMD{ J.~Teschner, ``Remarks on Liouville theory with
boundary,''{\tt hep-th/0009138}.
}

\lref\SeibergNM{ N.~Seiberg and D.~Shih, ``Branes, rings and
matrix models in minimal (super)string theory,'' {\it JHEP} {\bf 02}
(2004) 021,  {\tt hep-th/0312170}.
}

\lref\KutasovFG{ D.~Kutasov, K.~Okuyama, J.~Park, N.~Seiberg
and D.~Shih, ``Annulus amplitudes and ZZ branes in minimal string
theory,'' {\it JHEP} {\bf 08}  (2004) 026, {\tt hep-th/0406030}.
}

\lref\MaldacenaSN{ J.~Maldacena, G.~W.~Moore, N.~Seiberg and
D.~Shih, ``Exact vs. semiclassical target space of the minimal
string,'' {\tt hep-th/0408039}.
}

\lref\AganagicQJ{
M.~Aganagic, R.~Dijkgraaf, A.~Klemm, M.~Marino and C.~Vafa,
``Topological strings and integrable hierarchies,''
{\tt hep-th/0312085}.
}

\lref\WittenHR{ E.~Witten, ``Two-dimensional gravity and
intersection theory on moduli space,'' {\it Surveys Diff.\ Geom.\  }{\bf
1} (1991) 243.
}

\lref\GaiottoYB{ D.~Gaiotto and L.~Rastelli,
``A paradigm of open/closed duality: Liouville D-branes and the Kontsevich model,''
{\tt hep-th/0312196}.
}

\lref\KontsevichTI{ M.~Kontsevich,
``Intersection theory on the moduli space of curves and the matrix Airy function,''
{\it Commun.\ Math.\ Phys.\ } {\bf 147}  (1992) 1.
}

\lref\WittenMK{
E.~Witten,
``The $N$ matrix model and gauged WZW models,''
{\it Nucl.\ Phys. }  {\bf B371} (1992) 191.
}

\lref\DaulBG{ J.~M.~Daul, V.~A.~Kazakov and I.~K.~Kostov,
``Rational theories of 2D gravity from the two matrix model,''
{\it Nucl.\ Phys.\ } {\bf B409}  (1993) 311, {\tt hep-th/9303093}.
}

\lref\BanksDF{
T.~Banks, M.~R.~Douglas, N.~Seiberg and S.~H.~Shenker,
``Microscopic and macroscopic loops in nonperturbative two-dimensional gravity,''
{\it Phys.\ Lett.\ } {\bf B238} (1990) 279.
}

\lref\MooreIR{ G.~W.~Moore, N.~Seiberg and M.~Staudacher, ``From
loops to states in 2D quantum gravity,'' {\it Nucl.\ Phys.\ } {\bf
B362} (1991) 665.
}

\lref\MooreAG{
G.~W.~Moore and N.~Seiberg,
``From loops to fields in 2D quantum gravity,''
{\it Int. J.\ Mod.\ Phys.\ } {\bf A7} (1992) 2601.
}

\lref\DiFrancescoNW{ P.~Di Francesco, P.~H.~Ginsparg and
J.~Zinn-Justin, ``2D Gravity and random matrices,'' {\it Phys.\ Rept.\ }
{\bf 254} (1995) 1, {\tt hep-th/9306153}.
}

\lref\MorozovHH{ A.~Morozov, ``Integrability and matrix models,''
{\it Phys.\ Usp.\  } {\bf 37}  (1994) 1, {\tt hep-th/9303139}.
}

\lref\McGreevyKB{ J.~McGreevy and H.~Verlinde,
``Strings from tachyons: The $c = 1$ matrix reloaded,''
{\tt hep-th/0304224}.
}

\lref\KlebanovKM{ I.~R.~Klebanov, J.~Maldacena and N.~Seiberg,
``D-brane decay in two-dimensional string theory,''
{\it JHEP} {\bf 07}  (2003) 045, {\tt hep-th/0305159}.
}

\lref\Berry{ M.~V.~Berry,
``Stokes' phenomenon; smoothing a Victorian discontinuity,''
{\it Inst. Hautes \'Etudes Sci.\ Publ.\ Math.\ } {\bf 68} (1988) 211.
}

\lref\FidkowskiNF{
L.~Fidkowski, V.~Hubeny, M.~Kleban and S.~Shenker,
``The black hole singularity in AdS/CFT,''
{\it JHEP} {\bf 02}  (2004) 014, {\tt hep-th/0306170}.
}

\lref\GinspargIS{ P.~Ginsparg and G.~W.~Moore, ``Lectures on 2D
gravity and 2D string theory,'' {\tt hep-th/9304011}.
}

\lref\HarishChandra{
Harish-Chandra, ``Differential operators on a
semisimple Lie algebra,'' {\it Am. J. Math.} {\bf 79} (1957) 87
}

\lref\ItzyksonFI{
C.~Itzykson and J.~B.~Zuber,
``The planar approximation. 2,''
{\it J.\ Math.\ Phys.\ } {\bf 21} (1980) 411.
}

\lref\GiustoMT{
S.~Giusto and C.~Imbimbo,
``The Kontsevich connection on the moduli space of FZZT Liouville branes,''
{\tt hep-th/0408216}.
}

\lref\HashimotoSM{
A.~Hashimoto and N.~Itzhaki,
``Observables of string field theory,''
{\it JHEP} {\bf 01}  (2002) 028,
{\tt hep-th/0111092}.
}

\lref\GaiottoJI{
D.~Gaiotto, L.~Rastelli, A.~Sen and B.~Zwiebach,
``Ghost structure and closed strings in vacuum string field theory,''
{\it Adv.\ Theor.\ Math.\ Phys.\ }  {\bf 6} (2003) 403,
{\tt hep-th/0111129}.
}

\lref\BaumgartlIY{
M.~Baumgartl, I.~Sachs and S.~L.~Shatashvili,
``Factorization conjecture and the open/closed string correspondence,''
{\tt hep-th/0412266}.
}

\lref\ShapiroGQ{
J.~A.~Shapiro and C.~B.~Thorn,
``BRST invariant transitions between closed and open strings,''
{\it Phys.\ Rev.\ } {\bf D36} (1987) 432.
}

\lref\ShapiroAC{
J.~A.~Shapiro and C.~B.~Thorn,
``Closed string - open string transitions and Witten's string field theory,''
{\it Phys.\ Lett. }  {\bf B194} (1987) 43.
}

\lref\ZwiebachBW{
B.~Zwiebach,
``Interpolating string field theories,''
{\it Mod.\ Phys.\ Lett.\ } {\bf A7} (1992) 1079,
{\tt hep-th/9202015}.
}

\lref\GopakumarKI{ R.~Gopakumar and C.~Vafa, ``On the gauge
theory/geometry correspondence,'' {\it Adv.\ Theor.\ Math.\ Phys.\ }
{\bf 3}  (1999) 1415, {\tt hep-th/9811131}.
}

\lref\WittenAG{
E.~Witten, ``Algebraic geometry associated with matrix models of two
dimensional gravity,'' {\it Topological Methods in Modern Mathematics}, (Stony Brook, NY, 1991),  235--269, Publish or Perish, Houston, TX, 1993.
}

\lref\JarvisQM{
T.~J.~Jarvis, T.~Kimura and A.~Vaintrob,
``Moduli spaces of higher spin curves and integrable hierarchies,''
{\it Compos.\ Math.} {\bf 126} (2001) 157,
{\tt math.ag/9905034.}
}

\lref\Po{A.~Polishchuk, ``Witten's top Chern class on the moduli space of higher spin curves,''
{\tt math.ag/0208112}.
}

\lref\AbJa{D.~Abramovich and T.~Jarvis, ``Moduli of twisted spin curves,''  {\it Proc.\ Amer.\ Math.\ Soc.\ }
{\bf 131}  (2003)  685, {\tt math.ag/0104154}.
}

\lref\FaJaRu{H.~Fan, T.~Jarvis, and Y.~Ruan, ``Geometry and analysis of spin equations,''
{\tt math.dg/0409434}.
}

\lref\ItzyksonYA{ C.~Itzykson and J.~B.~Zuber, ``Combinatorics of
the modular group. 2. The Kontsevich integrals,'' {\it Int.\ J.\ Mod.\
Phys.\ } {\bf A7}  (1992) 5661, {\tt hep-th/9201001}.
}

\lref\KristjansenHN{ C.~Kristjansen, ``Multiloop correlators for
rational theories of 2D gravity from the generalized Kontsevich
models,'' {\it Nucl.\ Phys.\ } {\bf B436} (1995) 342,
{\tt hep-th/9409066}.
}

\writedefs

\newbox\tmpbox\setbox\tmpbox\hbox{\abstractfont }
\Title{\vbox{\baselineskip12pt \hbox{hep-th/0501141}
\hbox{MAD-TH-04-014} \hbox{PUPT-2149} \hbox{NSF-KITP-04-132}}}
{\vbox{\centerline{Open/Closed String Duality
}\smallskip \centerline{for Topological Gravity with Matter}}}
\smallskip
\centerline{Akikazu Hashimoto,$^a$ Min-xin Huang,$^a$ Albrecht
Klemm,$^a$ and David Shih$^b$}
\smallskip
\bigskip
\centerline{$^a${\it Department of Physics, University of
Wisconsin, Madison, WI 53706, USA}}
\medskip
\centerline{$^b${\it Department of Physics, Princeton University,
Princeton, NJ 08544, USA}}
\bigskip
\vskip 1cm

\noindent
The exact FZZT brane partition function for topological gravity
with matter is computed using the dual two-matrix model. We show
how the effective theory of open strings on a stack of FZZT branes
is described by the generalized Kontsevich matrix integral,
extending the earlier result for pure topological gravity. Using
the well-known relation between the Kontsevich integral and a
certain shift in the closed-string background, we conclude that
these models exhibit open/closed string duality explicitly. Just
as in pure topological gravity, the unphysical sheets of the
classical FZZT moduli space are eliminated in the exact answer.
Instead, they contribute small, nonperturbative corrections to the
exact answer through Stokes' phenomenon.

\vskip .5cm \Date{January 2005}



\newsec{Introduction}

Open/closed string duality is one of the most intriguing facets of
string theory. It states that dynamical processes involving open
strings can be formulated strictly in terms of closed strings, and
vice versa. Often, the open string side of the correspondence can
be reduced to an ordinary gauge theory. In such cases, the
correspondence can provide a promising way to address outstanding
issues of quantum gravity by reformulating them as questions in
gauge theory.

Recently, there has been renewed interest in ``minimal string
theory," i.e.\ non-critical strings with $c < 1$. (For a list of
references to recent work, see e.g.\ \SeibergAT.) These models are
dual to certain zero-dimensional gauge theories (i.e.\ matrix
models), providing a tantalizing hint that some form of
open/closed string duality is at work here. Because these systems
are integrable, it is natural to expect that $c< 1$ string
theories exhibit open/closed string dualities which are explicitly
demonstrable. This strongly motivates us to study these models further.

One of the main ingredients in formulating an open/closed string
duality are D-branes.  The D-branes relevant to non-critical
strings with $c < 1$ are called FZZT branes, and they were studied
in the context of Liouville theory in
\refs{\FateevIK,\TeschnerMD}. In the recent papers of
\refs{\SeibergNM\KutasovFG-\MaldacenaSN}, FZZT branes were used to
provide a geometric interpretation of minimal string theory, as
well as to probe the structure of its target space. Meanwhile,
they were used in \AganagicQJ\ to explore the relation of minimal
strings to topological strings on non-compact Calabi-Yaus with
B-branes.

Since our goal here is to understand open/closed string duality in
minimal string theory, it is natural to consider the dynamics
governing the open strings ending on FZZT branes. Exactly this
question was addressed for the case of the $c=-2$ model -- known as
pure topological gravity from the work of \WittenHR\ -- first from the
point of view of string field theory in \GaiottoYB, and later using
the double-scaled matrix model in \MaldacenaSN. Both approaches found
that the open strings ending on FZZT branes in this background are
described by the matrix integral of Kontsevich
\KontsevichTI. By identifying the precise deformation of the
closed string background associated with the presence of the
brane, the authors of \refs{\MaldacenaSN,\GaiottoYB} showed
concretely how Kontsevich's formulation of two dimensional gravity
can be thought of as a kind of open/closed string duality.

Pure topological gravity is often referred to as the $(2,1)$
model, by analogy with $(p,q)$ minimal string theory. It has
natural generalizations, commonly referred to as the $(p,1)$
models, which describe topological matter coupled to gravity. The
appropriate generalizations of the Kontsevich matrix integral also
exist \refs{\KontsevichTI,\WittenMK}. In this article, we will use
the double-scaled two-matrix model \DaulBG\ to relate the two
descriptions via open/closed string duality. That is, we will show
explicitly that the generalized Kontsevich integral describes the
effective theory of open strings between FZZT branes in the
$(p,1)$ closed-string background. Just as for the $(2,1)$ model,
we will see that the generalized Kontsevich integral includes
nonperturbative effects that drastically modify the topology of
the FZZT moduli space.

This paper is organized as follows. We begin in section 2 by
formulating the $(p,1)$ minimal string theories as a
double-scaling limit of the two matrix model. We compute the
partition function of a stack of FZZT branes in this background
and obtain the generalized (matrix) Airy function. We also explore
the Stokes and anti-Stokes lines for the FZZT partition function.
In section 3, we use the correspondence between macroscopic loop
operators and the local closed string operators to formulate the
open/closed string duality for the $(p,1)$ model. Finally, we
conclude in section 4 with a discussion of open problems and
relation to other work. The appendices contain various
generalizations and technical details.

\newsec{Double-scaling the $(p,1)$ models}

\subsec{FZZT branes in the two-matrix model}

In this section, we describe the analysis of the FZZT partition
function for general $(p,1)$ models using the two-matrix model,
extending the results of \MaldacenaSN. Let us begin by considering
a general form of the two-matrix model
\eqn\twomatdef{
\CZ(g) = \int dA\,dB\,e^{-{1\over g}\Tr(V(A)+W(B)-AB)} \ ,
}
where $A$ and $B$ are $N\times N$ matrices, $g$ is the coupling
constant of the bulk theory, and the choice of integration contour
depends on the form of the potentials $V$ and $W$. This model can
describe $(p,q)$ minimal string theory in the large $N$ limit
\DaulBG, provided we tune the potentials $V(A)$ and $W(B)$ to
\eqn\dkkVW{\eqalign{
V'(a) &= {1 \over 2 \pi i} \oint {Y_*(1/z) \over a - X_*(z)}
dX_*(z) \cr W'(b) &= {1 \over 2 \pi i} \oint {X_*(1/z) \over b -
Y_*(z)} dY_*(z)  \ ,
}}
where we integrate $z$ on a small contour around $z=0$, and
$X_*(z)$ and $Y_*(z)$ have zeros of order $q$ and $p$,
respectively, at $z=1$. We find it convenient to take
\eqn\dkkXY{\eqalign{
X_*(z) &= {(z-1)^q \over z} \cr
Y_*(z) &= {(1-z)^p \over z} \ .
}}
Then, for $q=1$, the potentials $V(a)$ and $W(b)$ take the simple
form
\eqn\pcritpot{
V_p(a) = -\sum_{k=1}^{p}{a^{k}\over k}+H_p,\qquad W_p(b) =  b \ .
}
Here we have fixed the integration constant to be
\eqn\intconst{
H_p=\sum_{k=1}^p {1\over k} } so that $V_p(1)=0$. This will be
convenient for later calculations. One can further simplify the
potential by shifting $A \rightarrow A+1$, which leads to
\eqn\twomattop{
\CZ_{(p,1)}(g) = \int dA\,dB\,e^{-{1\over g}\Tr(V_p(A+1)-AB)} \ . 
}
In the large $N$ double-scaling limit, this becomes the bulk
partition function of the $(p,1)$ model. Two comments on this
result are in order:

\lfm{1.} For $p=2$, one can integrate out $A$ to obtain a Gaussian
model for $B$. This case was studied in detail in \MaldacenaSN. Of
course, $A$ cannot be integrated out so easily in general.
Nevertheless, the integral \twomattop\ is essentially trivial for
any value of $p$, since $B$ always acts like a Lagrange multiplier
constraining $A$. This will facilitate much of the computation
reported in this article.

\lfm{2.} Some care is necessary in order to ensure that the matrix
integral in \twomattop\ is well defined. For even $p$, this can be
accomplished by integrating $A$ and $B$ with respect to the
measure where $\eta A$ and $i \eta^{-1} B$ are Hermitian matrices,
and $\eta^p = -1$. For odd $p$, one can use the same measure, but
one should integrate first over $B$ so that the $A$ integral is
constrained.

\medskip

We will consider exact D-brane observables constructed from the
macroscopic loop operator \refs{\BanksDF\MooreIR-\MooreAG}
\eqn\macroloop{
W(y)=\Tr \log(y-B) \ . 
}
Here $y$ is called the boundary cosmological constant, and it
parametrizes the moduli space of the FZZT brane. Following
\MaldacenaSN, the simplest such observable is the exact partition
function of the FZZT brane:
\eqn\detB{
\left\langle e^{W(y)} \right\rangle = \big\langle \det(y-B)
\big\rangle = \CZ_{p,1}(g)^{-1} \int dA\,dB\,e^{-{1\over
g}\Tr(V_p(A+1)-AB)} \det(y-B) \ .
}
Here the integral over $A$ and $B$ is defined in the same way as
in \twomattop, as discussed above. For the rest of this section,
we will focus mainly on the single determinant expectation value
\detB, since this captures the essentials of the matrix model
analysis. In section 2.4 we will also extend our analysis to the
correlator of multiple determinants.

The expectation value of $\det(y-B)$ can be computed efficiently
using the method of orthogonal polynomials (for a review and
original references, see e.g.\ \refs{\DiFrancescoNW,\MorozovHH}).
The bi-orthogonal polynomials are defined with the normalization
$P_m(a)=a^m+\dots$ and $Q_n(b)=b^n+\dots$, and satisfy the
orthogonality relation
\eqn\biorthdef{
\int da\,db\,e^{-{1\over g}(V_p(a+1)-ab)}P_m(a)Q_n(b) =
h_{m}\delta_{m,n} \ ,
}
where again, the integration over $a$ and $b$ is defined as in
\twomattop. For the orthogonality relation \biorthdef, it is
possible to write down explicit, closed-form expressions for $P_n$
and $Q_n$:
\eqn\biorth{\eqalign{
P_n(a) &= a^n\cr Q_n(b) &= \left. \left(-g{\partial\over\partial
z}\right)^n e^{{1\over g}(V_p(z+1)-bz)}\right|_{z=0}\ .
} }
It is straightforward to verify that  \biorth\ satisfies
\biorthdef.

In terms of these orthogonal polynomials, the expectation value of the
determinant is simply
\eqn\dettopii{
\langle \det(y-B)\rangle = Q_N(y) = \left.
\left(-g{\partial\over\partial z}\right)^N e^{{1\over
g}(V_p(z+1)-yz)}\right|_{z=0}\ .
}
Now it only remains to extract its double-scaling limit.

Before proceeding to this, it is instructive to re-derive the
result \dettopii\ using auxiliary fermions, as was done for the
$(2,1)$ model in \MaldacenaSN. Let us introduce fermionic
variables $\chi_i$ and $\chi^\dagger_i$ where $i$ runs from 1 to
$N$, and write
\eqn\detrew{
\det(y-B)=\int d\chi\,d\chi^\dagger e^{\chi^\dagger(y-B)\chi} \ .
}
Then, we can integrate out the matrices $A$ and $B$ as follows:
\eqn\dettopferm{\eqalign{
\langle \det(y-B)\rangle &= \CZ_{(p,1)}(g)^{-1}\int
dA\,dB\,d\chi\,d\chi^\dagger e^{-{1\over
g}\Tr(V_p(A+1)-AB)+\chi^\dagger(y-B)\chi}\cr &=\CZ_{(p,1)}(g)^{-1}
(2 \pi g)^{N^2}\int dA\,d\chi\,d\chi^\dagger e^{-{1\over g}\Tr
V_p(A+1)+y\chi^\dagger\chi}\delta(A+ g \chi\chi^\dagger)\cr &=\int
d\chi\,d\chi^\dagger e^{{1\over g}V_p(-
g\chi^\dagger\chi+1)+y\chi^\dagger\chi} \ . 
} }
In the last line we have used the fact that
$\Tr(\chi\chi^\dagger)^k=-(\chi^\dagger\chi)^k$ and the fact that the
partition function evaluates to
\eqn\zg{\CZ_{p,1}(g) = (2 \pi g)^{N^2} \ . }
The next step is to introduce auxiliary parameters $s$ and $z$ and
write
\eqn\dettopint{\eqalign{
 \langle \det(y-B)\rangle&= \int {ds\,  dz \over 2 \pi g} \int
 d\chi\,d\chi^\dagger\, e^{{1\over
 g}(V_p(z+1)-yz)}e^{{is\over g}(z+g\chi^\dagger\chi)}\cr
 &= \int {ds\,  dz \over 2 \pi g}\, e^{{1\over
 g}(V_p(z+1)- y z + isz)}(is)^N \ . 
 }}
By integrating $z$ by parts, the factor of $e^{isz/g}(is)^N$ can
be converted to $e^{isz/g}(-g
\partial_z)^N$ acting on the rest of the integrand.  The integral
over $s$ will then give rise to a delta function constraint
$\delta(z)$, and integrating over $z$ reproduces the answer
\dettopii\ for the expectation value of the determinant.

The derivation in terms of the fermions leads to the following
physical interpretation. The matrices $A$ and $B$ describe the
open strings between $N$ condensed ZZ branes. These provide the
open-string dual to the $(p,1)$ closed-string background, as shown
(for $c=1$) in \refs{\McGreevyKB,\KlebanovKM}. Introducing the
FZZT brane leads to $\chi$ and $\chi^\dagger$, which represent the
open strings (including orientation) stretched between the FZZT
brane and the condensed ZZ branes. Finally $s$ and $z$ represent
the degrees of freedom for the effective theory on the FZZT brane,
obtained after integrating out all the background degrees of
freedom. It is interesting that $s$ and $z$ seem to be conjugate
to one another (as do $A$ and $B$). In the next subsection, we
will take the double-scaling limit of the $s$, $z$ integral
\dettopint, and we will show that it reduces to the generalized
Airy function.

\subsec{Double-scaling Limit}

In order to take the double-scaling limit, it is convenient to write \dettopii\ in yet another integral form
\eqn\intform{
\langle \det(y-B)\rangle = (-g)^N N!   {1 \over 2 \pi i}  \oint
{dz \over z}  e^{{1 \over g} (V_p(z  +1) -yz )- N \log z    } \ , 
}
where the $z$ contour integral picks up the residue of the pole at
$z=0$. The  integrand contains $p$ saddle points in the complex
$z$ plane, located at the solutions of
\eqn\saddlez {(z+1)^p + yz + gN - 1 = 0\ . }
The saddles collide when $y \rightarrow 0$ and $gN =1$, and this
is the critical behavior that gives rise to the double-scaling
limit of the $(p,1)$ model \DaulBG. To extract the behavior of
\intform\ in the double-scaling limit, we set
\eqn\gNrel{
g = {1\over N}\ , 
}
and scale
\eqn\dblscllim{
 N = \epsilon^{-(p+1)}, \qquad z = -1 + \epsilon  \tilde z, \qquad y = \epsilon^p  \tilde y \ , }
while sending $\epsilon$ to zero.  We also deform the small
contour around $z=0$ to a contour $\CC$ along rays emanating from
$z=-1$:
\eqn\spzcontour{\CC: \quad \tilde z = \left\{ \eqalign{
&\hbox to 1in{$ -e^{-i  \pi/2(p+1)} t$, \hfil} t < 0 \cr &\hbox to
1in{$e^{i \pi/2(p+1)} t,$ \hfil }  t > 0\ . }\right.
 }
To see that this contour deformation is valid, notice that the
integrand of (2.16) goes to zero exponentially fast for $-{
\pi\over 2(p+1)} < \arg  \tilde z< { \pi\over 2(p+1)}$ and
$|\tilde z|\to \infty$. Thus, we can close the contour \spzcontour\ at
infinity without changing the value of the integral. This closed
contour encircles $z=0$, and since the integrand of \intform\ has
no other poles in the finite $z$ plane, the contour simply picks
up the residue at $z=0$. This shows that \spzcontour\ is indeed
equivalent to the original contour around $z=0$.

In the small $\epsilon$ limit, \intform\ becomes
\eqn\dslim{
\langle \det(y-B) \rangle =  {\N(y)} \int_{\cal C} {d \tilde
z\over 2 \pi i} \, e^{{\tilde z^{p+1} \over p+1} - \tilde y \tilde
z} = \N(y) Ai_p(\tilde y) \ , 
}
where we have defined
\eqn\pAidef{ Ai_p(\tilde y) = \int_{\cal C} {d \tilde z \over 2 \pi i}\,
e^{{  \tilde z^{p+1} \over p+1 } - \tilde y  \tilde z }\ , } and
$\N(y) =\sqrt{2 \pi N} N^{-1/(p+1)}  e^{-N(1-y)}$
is a non-universal overall factor. Removing this factor leads to
the final answer for the exact partition function $\psi(\tilde y)$
of the FZZT brane in the $(p,1)$ background:
\eqn\psidef{\psi(\tilde y) = \lim_{N \rightarrow \infty}
\N(y)^{-1}
 \langle \det (y - B)\rangle  =  Ai_p(\tilde y) \ . }
This is the generalization of the result of \MaldacenaSN\ to
higher $p$.

In order to avoid cluttering the equations, we will henceforth
drop the tildes from the double-scaled variables $y$ and $z$,
\eqn\notchange{
\tilde y \to y,\qquad \tilde z\to z \ . 
}
Since we will focus almost entirely on continuum quantities from
this point on, we hope this notational change will not confuse the
reader.

One recognizes that $Ai_p(y)$ satisfies the differential equation
 \eqn\pAidifrel{
 \left(-{\partial \over \partial y}\right)^p Ai_p(y) -
 y Ai_p(y) = 0 \ . 
 }
Different choices of contours lead to different linear
combinations of the homogeneous solutions of \pAidifrel, but the
particular contour \spzcontour\ is special for the following
reason. It gives rise to the unique solution of \pAidifrel\ which
is real for real $y$ and decays without oscillating at large
positive $y$. (For $p=2$, this function is precisely the Airy
function $Ai(y)$.) It is highly nontrivial that for all $p$, this
physical solution of \pAidifrel\ emerges unambiguously from the
double-scaling limit of the two-matrix model. This is one of the
main results of our paper.

By taking $y$ to be large and positive, one recovers the
asymptotic behavior
 \eqn \psiasym{ \log \psi(y) \approx -{p \over p+1}
 y^{(p+1)/p}  -{p-1 \over2p} \log y
 }
which we recognize as the large $y$ asymptotics of the FZZT disk
and annulus amplitudes \refs{\SeibergNM,\KutasovFG}. Below, we
will see that this semiclassical approximation is valid not only
for large, positive $y$, but for {\it all} large $y$ except $y\to
-\infty$.

Let us also mention that, after taking the double-scaling limit,
the contour of integration defining $\psi(y)$ can be deformed
arbitrarily, so long as the contour does not pass through
``mountains" at infinity. By ``mountains," we mean asymptotic
regions where the integrand grows without bound or decays slower
than $|z|^{-1}$. In practice, this means that the deformed contour
must asymptote to rays $-e^{-i\theta}t$ and $e^{i\theta}t$ as
$t\to -\infty$ and $t\to +\infty$, respectively, with $\theta$
satisfying
\eqn\solongas{
{\pi \over 2 (p+1)} < \theta < {3\pi\over 2 (p+1)} \ . 
}
The integral will converge fastest along the path of steepest
descent $\CC'$, which corresponds to $\theta= {i \pi\over
p+1}$, i.e.
 \eqn\steepestcontour{
\CC':\quad z \to \cases{ -e^{-{i \pi\over p+1}} t, &
 $t \to -\infty$ \cr e^{{i \pi\over p+1}} t, & $t\to +\infty$ \ . }
 }
Note that the steepest descent contour is at the same time the
stationary phase contour. In the next subsection, we will use this
contour to study the role of Stokes' phenomenon in the generalized
Airy integral \pAidef.

\subsec{Asymptotic expansion and Stokes phenomenon}

Using the explicit integral representation \pAidef\ and the
saddle-point approximation, it is straightforward to analyze the
large $|y|$ asymptotics of the generalized Airy functions. Just as
in the case of $p=2$, we expect to find a rich structure of Stokes
lines and anti-Stokes lines. (A succinct account of Stokes'
phenomenon can be found in \Berry.)

The integral \pAidef\ has $p$ saddle points at
 \eqn\Aisaddle{ z = z_n(y) = |y|^{1\over p}e^{{i\over p}(\theta+2\pi n)}\ , \qquad n = 0,\dots, p-1\ , 
 }
where $y=|y|e^{i\theta}$. The value of the argument in the
exponential at $n$-th saddle point is simply
\eqn\Aisaddleact{ S_n \equiv {1\over p+1}z_n(y)^{p+1}-y\, z_n(y)
 =  {-{p \over p+1}} |y|^{{p+1} \over p} e^{i{p+1 \over p} (\theta + 2 \pi n)}
\ .  }
From the form of the contour \spzcontour, it is not difficult to
see that for large positive $y$ only the $n=0$ saddle contributes
to the integral. The other saddles are inaccessible due to large
``mountains'' in the integrand of \pAidef\ at infinity.

As one varies $y$ in the complex plane, however, other saddles can
begin to contribute. This happens when the steepest descent
contour $\CC'$ collides with the other saddle points. The set of
points in the complex $y$ plane where this happens form what are
called the Stokes lines. A necessary condition for this to happen
is for two saddle points $z_n$ and $z_m$ to satisfy
\eqn\Stokescond{
{\rm Im}\,S_n = {\rm Im}\, S_m
}
since this means that the two saddles can be connected by a
stationary phase contour. \Stokescond\ is not a sufficient
condition for Stokes phenomenon to occur, since we are only
interested the particular stationary phase contour $\CC'$
described in \steepestcontour. In general, though, we expect from
\Stokescond\ that the first Stokes line closest to the positive,
real $y$ axis occurs at
\eqn\stokefirst{ \theta = \pm {(p+2)\pi \over 2 (p+1)}\ . }

In addition to new saddle points contributing and disappearing,
one should allow for the possibility of a saddle other than $n=0$
dominating. This would happen only when
\eqn\antistokescond{
{\rm Re}\,S_n = {\rm Re}\,S_0
}
for some $n$. Using \Aisaddleact, this becomes
\eqn\antistoke{
\cos\left({p+1 \over p} (\theta + 2 \pi n)\right) = \cos\left({p+1
\over p} \theta  \right) \ , 
}
whose only solution is $\theta = \pi$. The locus of points in the
complex $y$ plane where the dominant saddle gets exchanged with
another saddle are known as the anti-Stokes lines. For the
generalized Airy functions, the anti-Stokes lines are found to
always lie along $\theta = \pi$. In other words, the asymptotic
form of the generalized Airy function is described everywhere by
the analytic continuation of the $n=0$ saddle point, except along
the negative $y$ axis. Physically, this means that the
semiclassical approximation, described in \psiasym, is valid for
all large $y$ except $y\to -\infty$.

It is useful to illustrate this behavior through a concrete
example. Consider the contours of fixed phase for $p=5$. For small
$\theta$, only the dominant $n=0$ saddle lies along the contour of
steepest descent. As one increases $\theta$ from zero to $\theta =
7 \pi/12$, this contour collides with the $n=4$ contour. So for
$\theta > 7\pi/12$, both saddles $n=0$ and $n=4$ contribute.
Continuing along in this way, one finds multiple saddles
contributing and disappearing, summarized in the table below.

\bigskip

{\vbox{\ninepoint{
$$
\vbox{\offinterlineskip\tabskip=0pt \halign{\strut \hfil ~$#$~
\hfil \vrule&  \hfil ~#~ \hfil \cr
\theta = \arg(y)  &  Contributing Saddle \cr \noalign{\hrule} 0 <
\theta < {7 \over 12} \pi &0 \cr {7 \over 12} \pi < \theta < {9
\over 12} \pi & 0,4 \cr {9  \over 12}\pi < \theta < {11  \over
12}\pi & 0,3,4 \cr {11 \over 12} \pi < \theta < \pi & 0,2,4 \cr
}}$$}}}

\noindent A similar result can be obtained for $-\pi < \theta <
0$. Finally, at $\theta=\pi$, the $n=4$ saddle replaces the $n=0$
saddle as the dominant saddle, while at $\theta = -\pi$, the $n=1$
saddle takes over as the dominant saddle. The structure of the
saddle points and their interplay with the steepest descent
contour is illustrated in figure 1. The same analysis can be
repeated for general $p$. (The $p=3$ case in particular was
studied in \FidkowskiNF\ in a rather different context.)

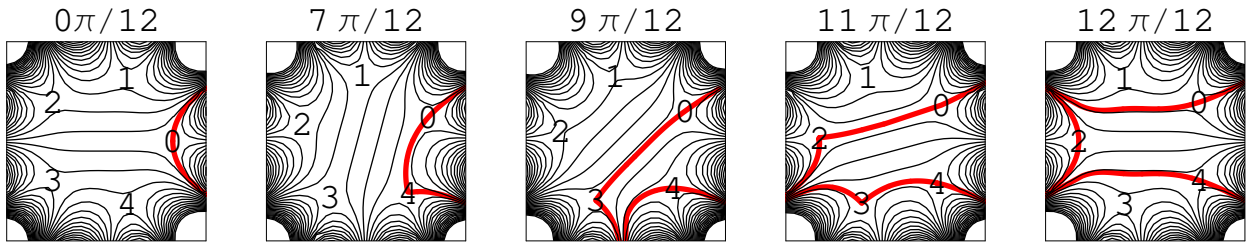
\medskip \ifig\figI{Contours of steepest descent in evaluating the integral expression \pAidef\ for the $p=5$ generalized Airy function for various $|y| \gg 1$  are illustrated in thick (red) line. ``0,1,2,3,4'' label the saddle points.  The phases of $y$ are chosen to coincide with the Stoke's lines.}
{\noindent \epsfxsize=\hsize\epsfbox{scontour.epsi}}

What is interesting about the structure of Stokes and anti-Stokes
lines for generalized Airy functions is that while the identity of
the dominant saddle is different between $\theta = \pi$ and
$\theta=-\pi$, the actual function is smooth along the anti-Stokes
line. The exact answer has no branch cut along the negative real
axis of the $y$ plane, despite the fact that the leading
perturbative amplitude \psiasym\ is a $p$-sheeted cover of the $y$
plane. Just as in the case of the $p=2$ model, the exact FZZT
amplitude is an entire function of $y$, and the unphysical sheets
of the FZZT moduli space are eliminated by non-perturbative
effects. However, as we see from the table above, the unphysical
sheets still have a physical effect: they contribute small,
non-perturbative corrections to the exact answer.

There is one sense in which the $p>2$ case is different from the
$p =2$ case.  The contribution from the dominant saddle
\Aisaddleact\ arbitrarily close to the negative real axis
\eqn\asymgenAi{S_0\to -{p \over p+1} |y| e^{\pm i {p +1 \over p} \pi}}
has positive real part for $p>2$. So $Ai_p(y)$ is oscillatory, but
it grows exponentially in magnitude as $y\to -\infty$. Contrast
this with the $p=2$ case, where it oscillates with an amplitude
which decays as a power of $y$. This difference between $p>2$ and
$p=2$ has the following matrix model interpretation. As was
discussed in \MaldacenaSN\ (see also section 9.3 of \GinspargIS),
the oscillatory power-law decay of $Ai(y)$ as $y\to -\infty$ is
directly related to the fact that the $p=2$ model can be described
by the algebraic eigenvalue density of a one-matrix model.
Applying the same reasoning to $p>2$ shows that these models {\it
cannot} be described by an algebraic density of a single
eigenvalue -- attempting to construct such an eigenvalue density
from $\psi(y)=Ai_p(y)$ would lead to a density that grows
exponentially as $y\to -\infty$. Of course, this is the expected
result since, as we saw above, these models are dual to two-matrix
models which cannot be reduced to one-matrix models.

\subsec{Multiple FZZT}

Finally, let us show how to extend the preceding analysis of the
single determinant expectation value to the case of multiple
determinants. Consider a stack of $m$ FZZT branes, whose boundary
cosmological constants are described by an $m\times m$ matrix $Y$
with eigenvalues $y_1,\dots,y_m$. The multiple FZZT partition
function in the matrix model is formulated as a double-scaling
limit of
\eqn\multidet{ \Big\langle
\det(Y\otimes I_N-I_m\otimes B) \Big\rangle = {\det_{ij}
Q_{N+i-1}(y_j) \over \Delta(y)}  \ , 
}
where $i$ and $j$ take on values $1,\dots,m$ and $\Delta(y)$ is
the Vandermonde determinant. The derivation of \multidet\ can be
found in e.g.\ \MorozovHH.

To evaluate this expression in the large $N$ limit, let us note that
according to \gNrel, a shift in $N$ is equivalent to a shift in
$g$. Now, $g$ is the coupling of the lowest dimension operator, and
for the $(p,1)$ models, a shift in this coupling is equivalent to a
shift in $y$. Therefore, \multidet\ becomes in the double-scaling
limit
\eqn\multidetbec{
 \Big\langle\det(Y\otimes I_N-I_m\otimes B) \Big\rangle \to
 \psi_m(Y) =
 {\det_{ij}\partial_{y_i}^{j-1}\psi(y_i)\over \Delta(y)} \ . 
 }
Here the finite $N$ and continuum $Y$ are related exactly as in
\dblscllim, and we have dropped irrelevant factors just as in
\psidef. This formula for the multi-FZZT correlator is a special
case of the general formula (applicable for all $(p,q)$ models) in
\MaldacenaSN.

Using \psidef\ along with the integral representation \pAidef, we
can turn \multidetbec\ into
\eqn\intmultiii{
\psi_m(Y) =  Ai_p(Y) \ , 
}
where $Ai_p(Y)$ is the generalized matrix Airy function
\eqn\matairy
{Ai_p(Y) =\Delta(y)^{-1}\int_{\cal C}  d z_i \, \Delta( z) e^{{
z_i^{p+1}  \over p+1} -  y_i  z_i} =  \int dZ\, e^{\Tr
\left({Z^{p+1} \over p+1} - YZ\right)} \ .
}
To establish the last equality, we used an identity from
\refs{\HarishChandra,\ItzyksonFI}. Eq.\ \intmultiii\ shows
explicitly how the effective theory on $m$ FZZT branes in the
$(p,1)$ background is described by the $m\times m$ generalized
Kontsevich integral.

\newsec{Open/closed string correspondence}

The main result of the previous section can be summarized as the
statement regarding the expectation value of the exponentiated
macroscopic loop operator
\eqn\macro{ \left\langle e^{W(y)} \right\rangle = \psi(y) = Ai_p(y)}
in the double-scaling limit. These macroscopic loop operators,
when parameterized by the length $\ell$ of the boundary on the
world sheet, can be decomposed into local {\it closed-string}
operators in terms of their scaling
\refs{\BanksDF\MooreIR-\MooreAG},
\eqn\scaling{ W(\ell) \sim \sum_{k\ge 1} {\ell^{k/p} \over \Gamma\left({k+p \over p}\right)} a_{p,k}   \sigma_k \ , }
where $a_{p,k}$ is a dimensionless normalization whose precise value
will be fixed in appendix B. The $\ell$ and the $ y$ parametrization
of the macroscopic loop operators are related by Laplace
transform
\eqn\laplace{ W( y) = \int {d\ell \over \ell} e^{-\ell  y} W(\ell) \sim \sum_{k\ge 1}{p \over k}  y^{-k/p} a_{p,k} \sigma_k \ . }
A similar relation between boundaries of fixed length and local
operators (as well as a Laplace transformed version) also appears
in section 3.1 of \KontsevichTI. One should keep in mind, however,
that the expansion of macroscopic loop amplitudes in terms of
microscopic operator correlation functions is subtle, in that
there are divergent contributions as $\ell=0$ or $ y \rightarrow
\infty$ in the disk and the annulus amplitudes \MooreIR. These can
be removed by introducing the normalization factor
\eqn\cyone{ C( y) =  {1 \over \sqrt{2 \pi} } e^{-{p \over p+1}  y^{(p+1)/p}  -{(p-1) \over2p} \log y} \ ,}
which precisely cancels the divergences from the disk and the annulus
amplitudes. One can then interpret the result of the previous section
as the statement
 \eqn\genfunc{
 \left\langle \exp\left({\sum_k {p \over k}  y^{-k/p} a_{p,k} \sigma_k }\right)
 \right\rangle = C( y)^{-1}\psi( y) = C( y)^{-1} Ai_p( y)\ , }
which relates the normalized FZZT brane partition function to the
generating function of closed-string correlators in $(p,1)$
topological string theory.

The relation \genfunc\ is the simplest statement of open/closed
string duality -- it says that the partition function of the FZZT
brane is equivalent to a certain shift in the closed-string
background. This relation becomes much richer if one considers
multiple FZZT branes, as in section 2.4. The left hand side of
\genfunc\ simply generalizes to
\eqn\genfuncii{
 \left\langle \exp\left({\sum_k {p \over k}
 \Tr \left(Y^{-k/p}\right) a_{p,k} \sigma_k }\right) \right\rangle \ , }
while the right hand side of \genfunc\ becomes
\eqn\Zclosed{
\Xi_m(Y) = C(Y)^{-1} Ai_p(Y) \ . }
Here $Ai_p(Y)$ is the matrix Airy function defined in \matairy,
and
\eqn\matrixAiasym{
C(Y) = \lim_{Y \rightarrow \infty}  Ai_p(Y)  =
  \int dZ\,
e^{\Tr \left( -{p \over 1+p}  Y^{(p+1)/p} + {1 \over 2} \sum_{i=0}^{p-1} Y^{i/p} Z Y^{(p-i-1)/p} Z \right)}}%
with the contour of $Z$ integration defined as in \matairy.
Combining all these ingredients, we arrive at the statement of
open/closed string duality
\eqn\genfunciii{
\left\langle \exp\left({\sum_k t_k \sigma_k }\right) \right\rangle
= \Xi_m(Y) , \qquad t_k = {p \over k} \Tr Y^{-k/p} a_{p,k}\ ,
}
which relates the generating function of topological closed string
amplitudes to the generalized Kontsevich integral.

Let us note that another way of writing \Zclosed, which may be
more familiar to some, comes from substituting $Y=X^p$
\eqn\ZclosedII{
\Xi_m(X) =   {\int dZ\,  e^{\Tr \left({Z^{p+1} -X^{p+1} \over p+1}
-(Z-X) X^p\right)} \over \int dZ\, e^{\Tr \sum_{i=0}^{p-1} {1
\over 2} X^i Z X^{p-i-1} Z}
}\ . }
This form of the generalized Kontsevich integral is common in the
literature (see e.g.\ \DiFrancescoNW), modulo trivial rescalings
and shifts of $Z$ and $X$. In the worldsheet (Liouville)
description of these models, $Y$ corresponds to the boundary
cosmological constants of the FZZT branes, and $X$ corresponds to
the dual boundary cosmological constants.

\newsec{Discussion}

In this article, we computed the exact partition function of
(multiple) FZZT branes in the $(p,1)$ topological background,
using the two-matrix model in the double-scaling limit. We found
that these partition functions are given by the generalized
Kontsevich integral. By relating this to a specific insertion of
closed string operators, we were able to formulate a precise
open/closed string duality for $(p,1)$ topological gravity with
matter.

In principle, the double-scaled matrix model can provide a
framework for understanding open/closed string duality in the most
general minimal string theory. However, one should keep in mind
that our analysis was aided by various simplifications that occur
in the $(p,1)$ models. The most important simplification came in
the calculation of the FZZT partition function at finite $N$. In
general, the FZZT partition function is the scaling limit of an
orthogonal polynomial of the dual matrix model. For the $(p,1)$
models, these orthogonal polynomials were sufficiently simple as
to admit an elementary, closed-form representation. Furthermore,
this representation allowed the scaling limit to be taken
explicitly, which led directly to the generalized Airy function.

It would be interesting to push the matrix model analysis of
open/closed string duality to more general $(p,q)$ models which
are non-perturbatively well defined. Of course, the orthogonal
polynomials for the general $(p,q)$ models are more complicated.
Nonetheless, they are known to satisfy a recursion relation and
can be generated in a finite number of steps. So there is still
hope that one may be able to formulate concretely open/closed
string duality for general $(p,q)$.

Another open problem of immediate interest is to rederive the results
in this paper using open-string field theory, as was done for $(2,1)$
in \GaiottoYB. (See also \GiustoMT\ for recent discussion on this
issue.) This would presumably shed more light both on open/closed
string duality in these models, and on the inner workings of
open-string field theory itself. It would also confirm the
identification of the generalized Kontsevich integral with the
effective theory of open strings between FZZT branes.

The primary motivation for studying explicit realizations of
open/closed string duality in toy systems like $(p,1)$ is to
provide new insights that can be extended to richer dynamical
systems. Let us briefly mention one possible lesson that could be
learned from our work. So far, all attempts to formulate purely
closed-string observables from open-string field theory in general
\refs{\HashimotoSM,\GaiottoJI} (see also the interesting recent
work of \BaumgartlIY), following the work of
\refs{\ShapiroGQ\ShapiroAC-\ZwiebachBW}, have had difficulty in
removing the boundary of the world sheet. By contrast, here we
encountered no such difficulties in formulating the open/closed
string duality of the $(p,1)$ model: purely closed-string
amplitudes were easily obtained from the generalized Kontsevich
matrix integral, and the latter presumably represents a reduction
of open-string field theory along the lines of \GaiottoYB.
Nonetheless, there was one seemingly unnatural step where we
removed the divergent contribution of the disk and the annulus
amplitudes by hand. Similar truncations of a few terms on the open
string side also arose in the context of Gopakumar-Vafa
correspondence \GopakumarKI.  It would be interesting if such
truncations are significant and have ramifications for open/closed
string dualities in general.

\vskip 0.8cm
\noindent {\bf Acknowledgments:}

\noindent We would like to thank
N.~Itzhaki, Y.~Ruan, and N.~Seiberg for useful discussions.  The work
of AH, MH, and AK are supported in part by DOE grant
DE-FG02-95ER40896.  MH was also supported by NSF grant PHY99-07949
during his stay at KITP. The research of DS is supported in part by an
NSF Graduate Research Fellowship and by NSF grant PHY-0243680. Any
opinions, findings, and conclusions or recommendations expressed in
this material are those of the author(s) and do not necessarily
reflect the views of the National Science Foundation.

\appendix{A}{Generalizing to other potentials}

In this article we studied the two matrix model with a single
multi-critical potential. However, it is very easy to generalize
this discussion to superpositions of multi-critical potentials.
This will allow us to use the two-matrix model to study the
integrable flows between multi-critical points with different
values of $p$, an aspect of minimal string theory that could not
be studied in \MaldacenaSN\ where the one-matrix model (which
always describes $p=2$) was used.

So let us consider the matrix integral
\eqn\VAsuperpos{
\CZ(g;s_i) = \int dA \, dB \,  e^{-  {1 \over g} (V(A) + W(B) -
AB)}\ , }
where the potential is determined using the formula \dkkVW\ of \DaulBG
\eqn\dkkXYgen{\eqalign{
X_*(z) &= {(z-1)^q \over z} \cr
Y_*(z) &= {(1-z)^p \over z} + \sum_{k=0}^{p-2} \epsilon^{-(p-k)} s_k {(1-z)^k \over z}\ .
}}
Then the expectation value of the determinant becomes
\eqn\dettopmixed{
\langle \det(y-B)\rangle \sim
    {1 \over 2 \pi i}  \oint  {dz \over z}  e^{{1 \over g} \left( V_p(z+1) + \sum_{k=0}^{p-2} (s_k \epsilon^{p-k} V_k(z +1)) - yz\right) - N \log z  }\ .}
One can then take the double-scaling limit
\eqn\dblsclimmixed{
N =  \epsilon^{-p-1},  \qquad y = \epsilon^{p}  \tilde y, \qquad z
= -1 + \epsilon  \tilde z, \qquad g N = 1 + \sum_{k=0}^{p-2}
\epsilon^{p-k} s_k}
Then as $\epsilon\to 0$, \dettopmixed\ becomes
\eqn\genairymixed{
 e^{-Ny} \langle\det(y-B)\rangle \sim
{1 \over 2 \pi i } \int_{\CC}d \tilde z\, e^{{ \tilde z^{p+1}
\over p+1} + \sum_{k=1}^{p-2}  s_k { \tilde z^{k+1} \over k+1}  -
\tilde y \tilde z} \equiv \psi(\tilde y)\ . }
where we have dropped numerical factors which do not depend on
$y$, and the contour are defined exactly as before.

One can show explicitly that the FZZT partition function
\genairymixed\ is the Baker-Akhiezer function of the KP hierarchy
with Lax pair
\eqn\QP{
Q = d^p+\sum_{k=0}^{p-1} s_k d^k\ ,\qquad P=Q^{1/p}_+=d\ ,
}
where $d= \partial_{s_0} = - \partial_{\tilde y}$. These clearly
satisfy
\eqn\QPcomm{
[P,Q]=1
}
and one can easily check that $\psi$ satisfies the equations
defining the Baker-Akhiezer function:
\eqn\psiQP{
Q\psi =  \tilde y\psi,\qquad P\psi=-\partial_{\tilde y} \psi \ . 
}
Note that we can convert this to a differential equation for
$\psi(\tilde y)$ as a function of $\tilde y$,
\eqn\genairyeq{
\psi^{(p)}(\tilde y)+\sum_{k=0}^{p-1} (-1)^{k} s_k \psi^{(k)}(\tilde y) = (-1)^p
\tilde y\, \psi(\tilde y)
}
which is a further generalization of  the  Airy equation.

\appendix{B}{Generalized Kontsevich integral and intersection numbers}

Here we will fix the normalization constants $a_{p,k}$ that arose
in the decomposition \scaling\ of the macroscopic loop operator
into closed-string scaling operators. Our method will be to match
$\Xi_m(X)$ given in \ZclosedII\ with the generating function
\eqn\taufunction{\tau(t)=\exp(F)=
\exp\left(\sum_{\{n,m\}} \langle \prod_{\{n,m\}}
\sigma_{n,m}^{d_{n,m}}\rangle \prod_{\{n,m\}}
{t_{n,m}^{d_{n,m}}\over d_{n,m}!}\right)} of intersection numbers
associated to $(p,1)$ topological string theory. These were
defined in \WittenAG\  as the integral
\eqn\defintersection{\left\langle\prod_{\{m_i,n_i\}} \sigma_{n_i,m_i}\right\rangle=
{1\over p^g}\int_{\overline {\cal M}_{g,s}(p, m)} \prod_{i=1}^s
c_1({\cal L}_i)^{n_i}  c_D({\cal V})\ } over the stable
compactification  of a covering $\overline {\cal M}_{g,s}(p,m)$ of
the moduli space $\overline {\cal M}_{g,s}$ of a Riemann surface
$\Sigma$ of genus $g$ with $s$ distinct marked points $x_i$,
$i=1,\ldots, s$. Here $\sigma_{0,m}$, $m=0,\ldots,p-2$ are the
primary matter fields and $n=0,\ldots$ labels the $n$-th
gravitational descendant $\sigma_{n,m}$. $\cal V$ is a vector
bundle over  ${\cal M}_{g,s}(p,{ m})$ with fiber $H^0(C,{\cal R})$
of dimension $D=(g-1)(1-{2\over p})+\sum_{i} {m_i\over p}$ and
$c_D({\cal V})$ is its top Chern class.   The line bundle ${\cal R}$
is defined up to isomorphism by ${\cal R}^{\otimes p}=K^{p-1}
\bigotimes_{i=1}^s {\cal O}(x_i)^{\otimes m_i}$, where $K$ is the
canonical bundle of $\Sigma$ and ${\cal O}(x_i)$ are line bundles
whose sections have at most simple poles at $x_i$. The covering
$\overline {\cal M}_{g,s}(p,m)$ depends on  data $m,p$ as
indicated by the notation.
$c_1({\cal L}_i)$ are the usual Mumford-Morita-Miller classes.
On dimensional grounds the integral above is zero unless
\eqn\vanishing{\sum_{i=1}^s(n_i+{m_i\over p}-1)=\left(3-\left(1-{2\over p}\right)\right)(g-1)\ .}
A mathematically precise construction of $c_D({\cal V})$ was
formulated in \refs{\JarvisQM,\Po}. The authors \JarvisQM\ furhter
confirmed the explicit calculation of intersection numbers in
\WittenAG. A suitable compactification of ${\cal M}_{g,s}(p, m)$ has
also been discussed in \refs{\AbJa,\FaJaRu}.

As was conjectured by \refs{\KontsevichTI,\WittenMK} and proven
for the case of $p=2$ in \KontsevichTI, \taufunction\ is the tau
function of the $p$ reduced KP hierarchy fulfilling in addition
the string equation. The geometric intersection numbers
\defintersection\ come with natural normalization, and the first few were worked out
in \WittenAG. We shall use \taufunction\ with this normalization
of the intersection numbers as the definition of the closed string
couplings $t_{n,m}$.

Following \ItzyksonYA, one can show that $F={\rm
lim}_{m\rightarrow \infty} \log(\Xi_m)$ fulfills the string
equation and the $p$ reduced KP hierarchy in the Miura variables
$t_k \sim {\rm Tr} (X^{-k})$. These properties determine $F$ up to
an additive constant. One can choose the normalization of $t_k$ so
that the string equation
\eqn\stringeq{{\partial F\over \partial t_{0,0}}=
{1\over 2} \sum_{i+j=p-2} t_{0,i} t_{0,j}+\sum_{n=0}^\infty
\sum_{m=0}^{p-2}t_{n+1,m} {\partial F\over \partial t_{n,m}}\ }
and the equations of the $p$ reduced KP hierarchy
\eqn\kp{{\partial^2 F\over \partial t_{0,0} \partial t_{n,k}}=
-c_{n,k} {\rm res}\left(Q^{n+{k+1\over p}}\right), \qquad Q = D^p
- \sum_{i=0}^{p-2} u_i(\{t_{n,m}\}) D^i, \qquad D  = {i  \over
\sqrt{p}} {\partial \over  \partial t_{0,0}}}
take on precisely the form used in \WittenAG, i.e.\ with
$c_{n,k}=(-1)^n p^{n+1}/\prod_{j=0}^n (j p+ k+1)$. This can be
achieved by setting\foot{In the case of $p=3$, the same change of
variables was described in \KristjansenHN.}
\eqn\variablechange{t_{np+k+1} = t_{n,k}=(-p)^{k-p-n(p+2)\over 2(p+1)}\prod_{j=0}^{n-1}
(jp+k+1) {\rm Tr}(X^{-(np+k+1)})\ . }
This fixes the normalization constant $a_{p,k}$ in \genfunciii.

As a check of our analysis, we have calculated a few of these
intersection numbers explicitly from the asymptotic expansion of
$\Xi_m(X)$, and compared them with results in the literature. To
extract the intersection numbers from $\Xi_m(X)$, we follow
\ItzyksonYA, the only difference being that the differential
operator $D$ with the defining properties
\eqn\diffopZ{
D \Xi_1 (x)={\int dz\, ze^{{z^{p+1} \over p+1}-z x^p+{p x^{p+1}
\over p+1}} \over \int dz\, e^{{p \over 2}x^{p-1}z^2}}\ ,
}
or $(D^p-x^p)\Xi_1(x)=0$, is given in our parametrization by
\eqn\diffop{D={p-1 \over 2p } {1 \over x^p}+x-{1 \over p x^{p-1}}
{d \over d x} \ . }
The expression for $\Xi_m(x)$ in terms of generalized characters
of the symmetric group in \ItzyksonYA\ gives the exact result for
all coefficients involving ${\rm Tr}(x^{-k})$ up to $k (p+1) \le
m$. Expanding in the corresponding order and using
\variablechange\ we obtain
\eqn\expresulta{\eqalign{F_{p=3}&=\left(
{1\over 2}t_{0, 0}^2 t_{0, 1}  + {1 \over 12} t_{1, 0}\right) \cr
&
 + \left({1\over 72} t_{0, 1}^4 +
          {1\over 2} t_{0, 0}^2 t_{0, 1} t_{1, 0} + {1\over 24} t_{1, 0}^2 +
          {1\over 6} t_{0, 0}^3 t_{1, 1} + {1\over 12} t_{0, 0} t_{2, 0}\right)+\ldots}}
and
\eqn\expresultb{\eqalign{F_{p=4} &=\left({1\over 2}t_{0, 0} t_{0, 1}^2 +
{1\over 2} t_{0, 0}^2 t_{0, 2} + {1\over 8} t_{1, 0}\right) 
+\biggl({1\over 16} t_{0, 1}^2 t_{0, 2}^2 +
        {1\over 2} t_{0, 0} t_{0, 1}^2 t_{1, 0} +
        {1\over 2} t_{0, 0}^2 t_{0, 2}  t_{1, 0} \cr &+ {1\over 16} t_{1, 0}^2 +
        {1\over 2} t_{0, 0}^2  t_{0, 1} t_{1, 1} + {1\over 6} t_{0, 0}^3 t_{1, 2} +
        {1\over 96} t_{0, 2} t_{1, 2} + {1\over 8} t_{0, 0} t_{2, 0}\biggr)+\ldots }
}
for $p=3$ and $p=4$ respectively.\foot{For $p=2$ our expansion
agrees with the one in \ItzyksonYA.} In particular, we find
agreement with the explicit calculations of \WittenAG\ for the
chiral ring correlators $\sigma_k=\sigma_{0,k}$
\eqn\expliciteintersections{
\eqalign{\langle \sigma_{k_1}\sigma_{k_2}\sigma_{k_3}\rangle&=
\delta_{k_1+k_2+k_3,p-2}\cr \langle
\sigma_{k_1}\sigma_{k_2}\sigma_{k_3} \sigma_{k_4}\rangle&= {1\over
p} {\rm min}(k_i,p-1-k_i)\ ,}}
from which all chiral ring correlators on the sphere follow by
associativity. We also agree with  $\langle
\tau_{1,0}\rangle=(p-1)/24$. More generally we find that for $p\le
10 $ and $k\le 3$, all predicted intersection numbers are positive
rational numbers. Further predictions for the  intersection
numbers with homogeneous degree $d$ where $\sigma_{n,k}$ has
degree $p n + k+1$ are tabulated below for $p=3,4,5$.

{\vbox{\ninepoint{
$$
{\vbox{\offinterlineskip
\tabskip=0pt
\halign{ \strut  \vrule#&  \hfil ~$#$~ \hfil & \vrule#&\hfil ~$#$~\hfil&\vrule#\cr
\noalign{\hrule}
&p=3 && d=12 &\cr
\noalign{\hrule}
&&&&\cr
&  \langle \sigma_{0,1}^4 \sigma_{1,0}\rangle_0  && {2\over 3}  & \cr
&  \langle \sigma_{0,0}^2 \sigma_{0,1}   \sigma_{1,0}^2 \rangle_0 && 2   & \cr
&  \langle \sigma_{0,0}^4 \sigma_{2,1}                 \rangle_0 && 1 & \cr
&  \langle \sigma_{0,0}   \sigma_{0,1}^3 \sigma_{1,1}\rangle_0 && {1\over 3} & \cr
&  \langle \sigma_{0,0}^3 \sigma_{0,1}   \sigma_{2,0} \rangle_0 && 1 & \cr
&  \langle \sigma_{0,0}^3 \sigma_{1,0}   \sigma_{1,1} \rangle_0 && 2 & \cr
&  \langle \sigma_{1,0}^3                           \rangle_1   && {1\over 6}   & \cr
&  \langle \sigma_{0,1}   \sigma_{1,1}^2            \rangle_1 && {1\over 36} & \cr
&  \langle \sigma_{0,0}   \sigma_{1,0}   \sigma_{2,0} \rangle_1 && {1\over 6} & \cr
&  \langle \sigma_{0,1}^2 \sigma_{2,1} \rangle_1 && {1\over 36} & \cr
&  \langle \sigma_{0,0}^2 \sigma_{3,0} \rangle_1 && {1\over 12} & \cr
&&&&\cr
\noalign{\hrule}
\multispan5 \phantom{XX}\cr
\multispan5 \phantom{XX}\cr
\noalign{\hrule}
&p=4 &&d=15 & \cr\noalign{\hrule}
&&&&\cr
&  \langle \sigma_{0, 0}^4 \sigma_{2, 2} \rangle_0  && 1  & \cr
&  \langle \sigma_{0, 0}^3 \sigma_{0, 1} \sigma_{2, 1} \rangle_0 && 1   & \cr
&  \langle \sigma_{0, 0}^3 \sigma_{0, 2} \sigma_{2, 0} \rangle_0 && 1 & \cr
&  \langle \sigma_{0, 0}^2\ \sigma_{0, 1}^2 \sigma_{2, 0} \rangle_0 &  & 1 & \cr
&  \langle \sigma_{0, 0}^3 \sigma_{1, 0} \sigma_{1, 2}  \rangle_0 && 2 & \cr
&  \langle \sigma_{0, 0} \sigma_{0, 1}^2\ \sigma_{0, 2} \sigma_{1, 2}  \rangle_0 &&
{1\over 4} & \cr
&&&&\cr
\noalign{\hrule}}}}\qquad
\vbox{\offinterlineskip\tabskip=0pt
\halign{ \strut \vrule#&&  \hfil ~$#$ \hfil & \vrule#&\hfil ~$#$\hfil~&
\vrule#\cr \noalign{\hrule}
&p=4 {\ \rm cont.}&&d=15 & \cr\noalign{\hrule}
&&&&\cr
&  \langle \sigma_{0, 1}^4 \sigma_{1, 2} \rangle_0   && {1\over 4}   & \cr
&  \langle \sigma_{0, 0}^3 \sigma_{1, 1}^2 \rangle_0 && 2 & \cr
&  \langle \sigma_{0, 0}^2 \sigma_{0, 1} \sigma_{1, 0} \sigma_{1, 1}\rangle_0 && 2 & \cr
&  \langle \sigma_{0, 0} \sigma_{0, 1} \sigma_{0, 2}^2 \sigma_{1, 1} \rangle_0  && {1\over 4} & \cr
&  \langle \sigma_{0, 1}^3 \sigma_{0, 2} \sigma_{1, 1}  \rangle_0 && {1\over 2} & \cr
&  \langle \sigma_{0, 0}^2\sigma_{0, 2}\sigma_{1, 0}^2  \rangle_0 && 2 & \cr
&  \langle \sigma_{0, 0}\sigma_{0, 1}^2 \sigma_{1, 0}^2 \rangle_0 && 2 & \cr
&  \langle \sigma_{0, 1}^2\sigma_{0, 2}^2 \sigma_{1, 0} \rangle_0 && {1\over 2} & \cr
&  \langle \sigma_{0, 2}^5 \rangle_0 && {1\over 8} & \cr
&  \langle \sigma_{0, 0} \sigma_{1, 2}^2 \rangle_1 && {1\over 48} & \cr
&  \langle \sigma_{0, 0}^2 \sigma_{3, 0} \rangle_1 && {1\over 8} & \cr
&  \langle \sigma_{0, 0} \sigma_{0, 2} \sigma_{2, 2} \rangle_1 && {1\over 96} & \cr
&  \langle \sigma_{0, 1}^2 \sigma_{2, 2},\rangle_1 && {1\over 32} & \cr
&  \langle \sigma_{0, 1} \sigma_{0, 2} \sigma_{2, 1} \rangle_1 && {1\over 24} & \cr
&  \langle \sigma_{0, 0} \sigma_{1, 0} \sigma_{2, 0}, \rangle_1 && {1\over 4} & \cr
&  \langle \sigma_{0, 2}^2 \sigma_{2, 0} \rangle_1 && {1\over 48} & \cr
&  \langle \sigma_{0, 1} \sigma_{1, 1} \sigma_{1, 2} \rangle_1 && {1\over 24} & \cr
&  \langle \sigma_{0, 2} \sigma_{1, 0} \sigma_{1, 2} \rangle_1 && {1\over 48} & \cr
&  \langle \sigma_{1, 0}^3   \rangle_1 && {1\over 4} & \cr
&  \langle \sigma_{0, 2} \sigma_{1, 1}^2 \rangle_1 && {1\over 24} & \cr
&  \langle \sigma_{3, 2} \rangle_2 && {3\over 2560} & \cr
&&&&\cr
\noalign{\hrule}}}\qquad
\vbox{\offinterlineskip\tabskip=0pt
\halign{ \strut \vrule#&&  \hfil ~$#$ \hfil & \vrule#&\hfil ~$#$\hfil~&
\vrule#\cr \noalign{\hrule}
&p=5 && d=6,12 &\cr\noalign{\hrule}
&&&&\cr
&  \langle  \sigma_{0,0}^2 \sigma_{0,3}\rangle_0 && 1  & \cr
&  \langle  \sigma_{0,0} \sigma_{0,1} \sigma_{0,2} \rangle_0 && 1  & \cr
&  \langle  \sigma_{0,1}^3 \rangle_0 && 1  & \cr
&  \langle  \sigma_{0,0}^3 \sigma_{1,3}\rangle_0 && 1  & \cr
&  \langle \sigma_{0, 0}^2 \sigma_{0, 1} \sigma_{1, 2}  \rangle_0 && 1   & \cr
&  \langle \sigma_{0, 0}^2 \sigma_{0, 2} \sigma_{1, 1}  \rangle_0 && 1 & \cr
&  \langle \sigma_{0, 0} \sigma_{0, 1}^2 \sigma_{1, 1} \rangle_0 && 1 & \cr
&  \langle \sigma_{0, 0}^2\ \sigma_{0, 3} \sigma_{1, 0}  \rangle_0 && 1 & \cr
&  \langle \sigma_{0, 0} \sigma_{0, 1} \sigma_{0, 2} \sigma_{1, 0}  \rangle_0 && 1 & \cr
&  \langle \sigma_{0, 1}^3 \sigma_{1, 0} \rangle_0   && 1   & \cr
&  \langle  \sigma_{0, 1}^2\ \sigma_{0, 3}^2\rangle_0 && {1\over 5} & \cr
&  \langle \sigma_{0, 1} \sigma_{0, 2}^2\sigma_{0, 3}  \rangle_0 && {1\over 5} & \cr
&  \langle \sigma_{0, 2}^4 \rangle_0 && {2\over 5} & \cr
&  \langle \sigma_{0, 0} \sigma_{2, 0} \rangle_1 && {1\over 6} & \cr
&  \langle \sigma_{0, 2}\ \sigma_{1, 3}, \rangle_1 && {1\over 60} & \cr
&  \langle  \sigma_{0, 3} \sigma_{1, 2}\rangle_1 && {1\over 60} & \cr
&  \langle  \sigma_{1, 0} \rangle_1 && {1\over 6} & \cr
&  \langle  \sigma_{1, 0}^2 \rangle_1 && {1\over 6} & \cr
&&&&\cr
&&&&\cr
&&&&\cr
&&&&\cr
\noalign{\hrule}}}
$$}}}

Finally, let us emphasize that, despite all the progress in making
Witten's conjecture mathematically precise
\refs{\JarvisQM\Po\AbJa-\FaJaRu}, a proof of \taufunction\ is
still lacking for $p>2$. That is, it remains to extend beyond
$p=2$ Kontsevich's statement that the intersection numbers for
$(p,1)$ topological gravity are generated by the tau function of
the $p$ reduced KP hierarchy. Such a generalization will
presumably involve open string field theory techniques in order to
generate the cell decomposition of the moduli space of spin
curves. See \refs{\GaiottoYB,\GiustoMT} for a recent discussion on
this issue.

\listrefs
\end